\documentclass[reprint,twocolumn,nobibnotes,amsmath,amssymb,aps,prl,floatfix]{revtex4-2}
\usepackage{hyperref}
\usepackage{url}
\DeclareUrlCommand\path{\urlstyle{same}}

\usepackage{xcolor}
\usepackage{graphicx}% Include figure files
\usepackage{dcolumn}% Align table columns on decimal point
\usepackage{bm}% bold math
\renewcommand\Re{{\rm Re}\,}
\newcommand\We{{\rm We}\,}

\begin{document}

\preprint{APS/123-QED}

\title{\textbf{Lacunarity in a Two-Dimensional Atomizing Liquid Jet} 
}% 

\author{Guangnian Ji}
\email[Contact author: ]{guangnian.ji@dalembert.upmc.fr}
\affiliation{Sorbonne Université and CNRS, UMR 7190, Institut Jean Le Rond $\partial$'Alembert, 75005 Paris, France}

\author{Yash Kulkarni}
\email[Contact author: ]{kulkarniyash2398@gmail.com}
\affiliation{Sorbonne Université and CNRS, UMR 7190, Institut Jean Le Rond $\partial$'Alembert, 75005 Paris, France}

\author{Stéphane Zaleski}
\email[Contact author: ]{stephane.zaleski@sorbonne-universite.fr}
\affiliation{Sorbonne Université and CNRS, UMR 7190, Institut Jean Le Rond $\partial$'Alembert, 75005 Paris, France}
\affiliation{Institut Universitaire de France}

\date{\today}% It is always \today, today,
             %  but any date may be explicitly specified

\begin{abstract}
Atomization stretches and folds the liquid--gas interface before fragmenting it into ligaments and droplets, making fractal measures a natural descriptor of the lacunarity of the final breakup state. We examine this idea in two-dimensional (2D) VOF-DNS of a liquid jet with adaptive mesh refinement in Basilisk. Box-counting of the full resolved interface does not yield a single scale-independent exponent. Instead, two scaling ranges appear, separated by a crossover near the box-counting level \(L_{\mathrm{box}}\simeq7\): coarser boxes measure the folded connected jet envelope, whereas finer boxes increasingly sample ligaments, droplets, and nearly smooth local interface segments. Decomposing the interface into detached droplets, ligaments, and the connected main body shows that the relevant effective dimension is structure dependent. Droplets are most lacunar with dimension below one, ligaments occupy an intermediate level, and the main body carries the largest coarse-scale dimension. This hierarchy persists over $Re_l=100$--$10^4$ at fixed $\We_g=200$. The lacunarity of droplets is reminiscent of that of the object positions in the night sky hypothesized by de Vaucouleurs. 
\end{abstract}

\maketitle

%Introduction

\clearpage

%\keywords{Suggested keywords}%Use showkeys class option if keyword
                              %display desired

Lacunarity is the almost-empty character of a statistically self-similar point set such as stars in the night sky \cite{smith1986lacunarity,de1970case}. It is characterized by the fractal dimension $D_F$ being smaller than the embedding space dimension $D$ as first discussed for turbulent fields by Mandelbrot\cite{mandelbrot1983fractal,sreenivasan1991fractals}. In turbulence, the set may be the region of strong dissipation, the boundary of the vortical region \cite{Sreenivasan_Meneveau_1986}, or a passively advected and initially smooth line of points. All these objects have dimension $1 < D_F < 2$ showing significant but not extreme lacunarity. In this work, we study the fractal dimension of the set of droplets, ligaments and the main body of an atomizing jet\cite{lefebvre2017atomization,bayvel2019liquid}, finding striking lacunarity ($D_F <1$) for the droplets.

%The underlying motivation is the statistical description of  multiphase turbulence.  In general in fully developed regimes the emergence of self-similarity is a fundamental problem of nonequilibrium physics \cite{richardson1922weather,kolmogorov1962refinement,kolmogorov1991local,sreenivasan1991fractals}. Jet atomization provides a particularly stringent multiphase realization of this problem \cite{lefebvre2017atomization,bayvel2019liquid}. A connected liquid mass is stretched, folded, and corrugated by a hierarchy of vortical motions before changing topology and fragmenting into ligaments and droplets. The geometrical object whose statistics are sought therefore evolves simultaneous{ly in space, scale, and topology. This raises the following questions: can such a breakup state exhibit a self-similar interfacial geometry, and over what range of scales is that description physically meaningful? How do the characteristic of multiphase turbulence compare to that of single phase turbulence? 

The core motivation of the fractal analysis is the investigation of the statistical properties of the flow. In a large body of numerical \cite{Kulkarni_Pairetti_Villiers_Popinet_Zaleski_2025} and experimental work \cite{villermaux2020fragmentation},  this analysis proceeds through the probability distribution function (PDF). Indeed atomization is commonly characterized through the number and size distribution of the droplets it produces, and thus by its PDF. 
However the latter primarily describes the detached fragmented population and does not quantify the connected jet, the attached ligaments, or the folded interface from which the droplets emerge. Consequently
a complementary statistic such as the fractal dimension is  needed to characterize the complete breakup geometry across scales.
To be specific, for a self-similar object made of $N$ copies of itself, each reduced by a factor $\epsilon$, the fractal dimension is $D=\log N/\log(1/\epsilon)$.
This leads to the box-counting definition of dimension, where $N(\ell)$ is the number
of boxes of size $\ell$ containing elements of the set. 
Then $N(\ell)\sim \ell^{-D}$, so $D$ measures how rapidly the number of spatial elements containing the interface grows as the observation scale is refined.

\begin{figure*}[t] 
\centering 
\includegraphics[width=\textwidth]{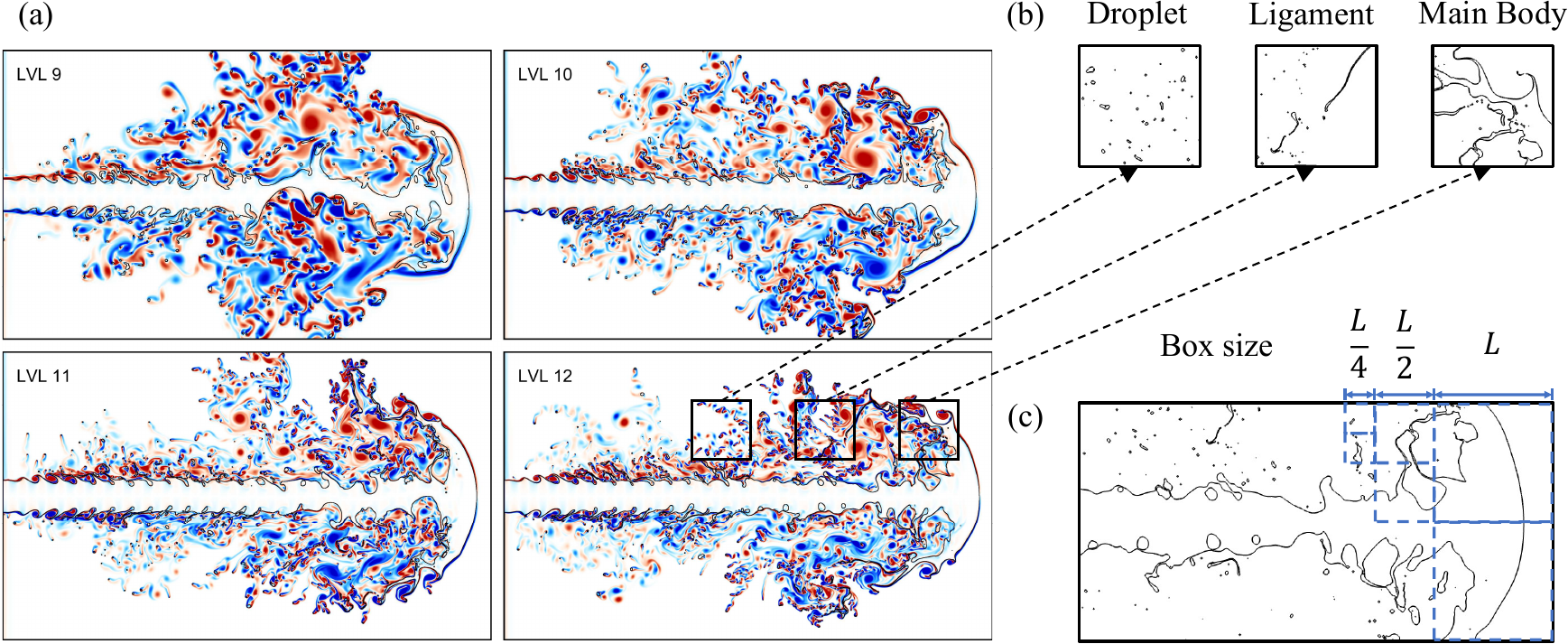} 
\caption{
(a) Interface and vorticity field at $t=3.0$ and $u_{e,\max}=10^{-6}$ for the reference case $\Re_l=5800$ and $\We_g=200$, shown for increasing maxlevel from $9$ to $12$. Higher refinement levels resolve smaller vortices and reveal finer interfacial corrugations, ligaments, and detached droplets.
(b) Representative local geometrical subsets extracted from the refined interface: droplets, ligaments, and the connected main body.
(c) Schematic of the box-counting procedure. A uniform grid is laid over the computational domain, and boxes intersecting the reconstructed interface are counted. Coarser boxes sample the large-scale jet envelope, whereas finer boxes probe local ligaments, droplets, and short interface segments.
} 
\label{fig:fig1} 
\end{figure*}

The fractal analysis of atomizing fields finds its roots in the analysis of turbulence, where the Richardson--Kolmogorov cascade picture describes the flow as a hierarchy of eddies transferring energy from large to small scales until the viscous cut-off \cite{richardson1922weather,kolmogorov1962refinement,kolmogorov1991local,sreenivasan1991fractals}.
Mandelbrot and others emphasized that such multiscale structures cannot be fully described by Euclidean geometry \cite{Sreenivasan_Meneveau_1986,mandelbrot1983fractal}.
The idea has been used to quantify interfaces extracted from experimental images or numerical fields, including isoscalar surfaces, turbulent--non-turbulent boundaries, and flame fronts \cite{prasad1990quantitative,constantin1991fractal,Xu_Long_Wang_2023,Soligo_Chiarini_Rosti_2025,gouldin1987application}.
Fractal measures have been performed in other multiscale flows \cite{el2022fractal,muller1992implication,constantin1985determining,heinen2015classical,el2023foam}.
Specifically, for atomizing jets scale similarity and fractal dimensions have been studied through analysis of experimental data \cite{shavit1994fractal,shavit1995fractal,castrejon2003fractal,chehroudi2004fractal,dumouchel2005experimental,grout2007fractal,le2008fractal,rimbert2021direct} but not in simulation except for the recent work of Herbert et al. \cite{Herbert2026}. The latter
use a regular grid discretization with the {\em ParisSimulator code} \cite{aniszewski2021parallel} while in what follows we use the quadtree capabilities of {\em Basilisk} \cite{basilisk}. Because of the close connection between quadtree adaptation and fractal dimension \cite{2027-popinet} our approach gives an interesting methodological alternative. Moreover in what follows we also distinguish among  droplets, ligaments and the bulk jet finding a separate scaling range and fractal dimension for each. 

In what follows we study a 2D version of the pulsed liquid jet configuration that was previously investigated by two of us  \cite{Kulkarni_Pairetti_Villiers_Popinet_Zaleski_2025}  and constitutes a paradigmatic numerical benchmark of atomizing flow.  The flow has Reynolds numbers $\Re_p = \rho_p U_I d_0 / \mu_p$
for phase $p$ with density $\rho_p$, viscosity $\mu_p$, injection velocity $U_I$ and jet diameter at injection $d_0$. The Weber numbers are $\We_p = \rho_p U_I^2 d_0/\sigma$ with $\sigma$ the surface tension.  Unlike the case of ref. \cite{Kulkarni_Pairetti_Villiers_Popinet_Zaleski_2025} we study this flow with a variable Reynolds number up to $\Re_l=10^4$. As in ref. \cite{Kulkarni_Pairetti_Villiers_Popinet_Zaleski_2025} we keep a constant Weber number $\We_g=200$. 
The 2D formulation rather than the 3D one in our previous work and that of Herbert et al.  \cite{Herbert2026} is chosen to access a wide scale range 
at moderate computational cost. 

{The incompressible Navier--Stokes equations are solved using the volume of fluid (VOF) method with quadtree adaptive mesh refinement in the Basilisk framework} \cite{basilisk,popinet2015quadtree}.
{The reference case has $\Re_l=5800$ and the} snapshot analyzed below is taken at $t=3.0$, with dynamic mesh refinement controlled by the residual bound $u_{e,\max}=10^{-6}$.
Figure~\ref{fig:fig1}(a) shows this reference case for maximum quadtree refinement level (``maxlevel'') in the range \(9\)--\(12\), illustrating that increasing resolution modifies the resolved vortical field and reveals progressively finer interfacial corrugations, ligaments, and droplets.
Because the calculation is 2D, the liquid--gas interface is a set of reconstructed curves and droplets appear as closed contours.
The dimensions reported below therefore quantify the accumulation of interfacial length across resolved scales, rather than the fractal dimension of a three-dimensional interfacial surface.

We use box counting as a finite-scale diagnostic in which the observation scale can be varied independently of the DNS resolution.
The key issue is whether a single exponent can characterize the fully resolved interface, or whether the measured dimension depends on both the observation scale and the geometrical subset being sampled.
The morphology in Fig.~\ref{fig:fig1} motivates separating the interface into droplets, ligaments, and the connected main body.
For a covering level \(L_{\mathrm{box}}\), the computational domain is tiled by a uniform grid of square boxes of size
\(\ell_{\mathrm{box}}=L_{\mathrm{dom}}/2^{L_{\mathrm{box}}}\), independent of the adaptive mesh used in the simulation.
We count the number \(N(L_{\mathrm{box}})\) of boxes intersecting the reconstructed liquid--gas interface, as sketched in Fig.~\ref{fig:fig1}(c).
For an ideal self-similar set, \(N(L_{\mathrm{box}})\sim \ell_{\mathrm{box}}^{-D}\), or equivalently \(N(L_{\mathrm{box}})\sim 2^{D L_{\mathrm{box}}}\) up to a constant factor.
The effective dimension is therefore obtained from the slope of \(\log N\) versus \(\log(1/\ell_{\mathrm{box}})\), equivalently from \(\log_2 N\) versus \(L_{\mathrm{box}}\).
The maxlevel determines which interfacial structures are resolved by the DNS, whereas \(L_{\mathrm{box}}\) defines the observation scale used to measure their geometry.
Before applying box counting to the atomizing interface, we assess its reliability and physical basis using a two-dimensional single-phase liquid jet (see Sec.~S1 of the Supplemental Material~\cite{SupplementalMaterial}).

\begin{figure}[!t]
\centering
\includegraphics[width=0.95\columnwidth]{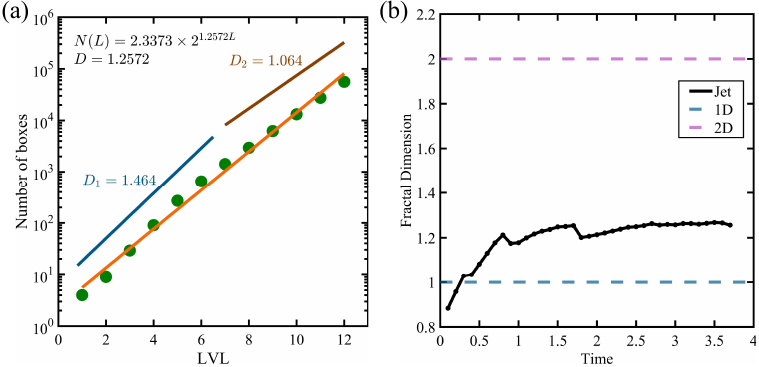}
\caption{
(a) Box count \(N(L_{\mathrm{box}})\) for the full reconstructed
liquid--gas interface of the reference case with maxlevel \(=12\).
The overall fit gives \(D\simeq1.257\), while separate ranges reveal
a coarse-scale slope \(D_1\simeq1.464\) and a fine-scale slope
\(D_2\simeq1.064\).
(b) Time evolution of the global effective dimension, compared with
the Euclidean values \(D=1\) and \(D=2\).
}
\label{fig:fig2}
\end{figure}

\textit{Full-interface scaling}--We first apply box-counting to the fully reconstructed interface of the reference case with maxlevel \(=12\).
This set contains the connected main body, ligaments, and detached droplets identified in Fig.~\ref{fig:fig1}(b).
Figure~\ref{fig:fig2}(a) shows that a fit over all covering levels gives an apparent global effective dimension \(D \simeq 1.257\).
This value is useful as a compact descriptor of the overall breakup state, but it should not be interpreted as a single scale-independent fractal dimension of the full interface.
Instead, the data exhibit a clear crossover.
At coarse covering levels, the larger slope \(D_1 \simeq 1.464\) reflects the large-scale, folded envelope of the connected jet, whereas at finer covering levels the slope decreases to \(D_2 \simeq 1.064\), consistent with the count becoming increasingly dominated by locally linear interface segments, ligaments, and droplet contours.

\begin{figure*}[!t]
\centering
\includegraphics[width=\textwidth]{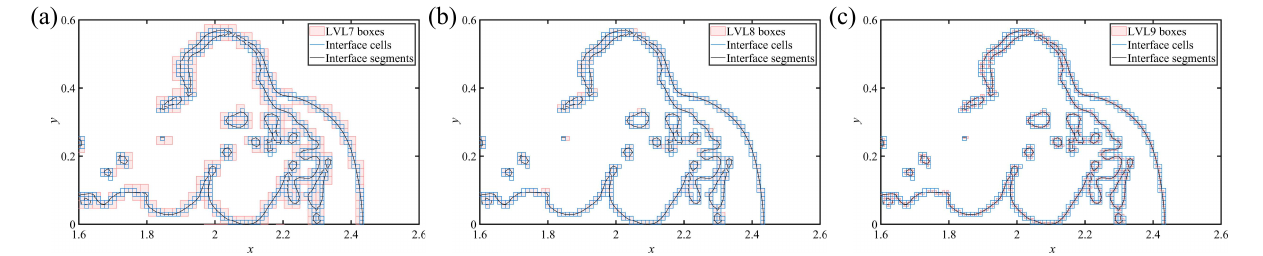}
\caption{
The same local interfacial region from a simulation with maxlevel
\(=8\), covered by box-counting grids with
(a) \(L_{\mathrm{box}}=7\),
(b) \(L_{\mathrm{box}}=8\), and
(c) \(L_{\mathrm{box}}=9\).
Blue boxes show the adaptive interfacial cells, black lines show the
reconstructed interface segments, and red boxes indicate the covering
boxes intersecting the interface.
The comparison illustrates how the observation scale changes the
geometry sampled by the count.
}
\label{fig:fig3}
\end{figure*}

The behavior at both ends of the covering-level range is expected. 
At very coarse \(L_{\mathrm{box}}\), the boxes are comparable to the jet envelope or even to the computational domain, so the count is affected by finite-size effects rather than by intrinsic interfacial scaling. 
At the opposite end, the covering-box size approaches the smallest resolved interfacial scale. 
In a geometrical VOF method, the liquid--gas interface is represented in interfacial cells by piecewise-linear reconstructed facets \cite{hirt1981volume,popinet2009accurate,basilisk}; in two dimensions, these facets are short line segments, as visualized in Fig.~\ref{fig:fig3}.
The box count then increasingly probes a locally rectifiable one-dimensional set, for which \(N(\ell_{\mathrm{box}})\sim \ell_{\mathrm{box}}^{-1}\). 
The effective slope therefore tends toward unity at the finest observable scales.
Such dependence on the fitting interval is expected in finite-resolution estimates of fractal dimension, where both finite-size effects and small-scale cutoffs can affect the apparent exponent \cite{theiler1990estimating,falconer2013fractal,feder2013fractals}.

Figure~\ref{fig:fig2}(b) shows that the global effective dimension, although only an apparent full-interface measure, becomes nearly stationary after the initial transient, fluctuating weakly around \(D\simeq1.25\). 
This does not by itself demonstrate exact self-similarity, but it does indicate that the fully developed jet reaches a statistically steady geometrical state.

The local coverings in Fig.~\ref{fig:fig3} illustrate the mechanism behind the crossover.
To visualize the effect of the covering level, a representative local interfacial region from a simulation with maxlevel \(=8\) is covered by box-counting grids with \(L_{\mathrm{box}}=7\), \(8\), and \(9\). 
As \(L_{\mathrm{box}}\) increases, the sampled geometry shifts from the large-scale jet envelope toward local reconstructed interface segments, ligaments, and droplet contours. 
The measured exponent is therefore controlled not only by the interfacial structures resolved by the maxlevel, but also by the observation scale \(L_{\mathrm{box}}\) used in the box-counting analysis. 

\textit{Structure-resolved hierarchy}--The crossover near \(L_{\mathrm{box}}\simeq7\) can be interpreted as a finite-scale geometrical transition rather than as a universal level.
In the present domain, it corresponds to a box size \(\ell_c=L_{\mathrm{dom}}/2^7\).
For \(\ell_{\mathrm{box}}>\ell_c\), the count is dominated by the large-scale folded envelope of the connected jet: shear-layer vortices stretch, corrugate, and fold the liquid interface, consistent with the intuitive view of turbulent interfacial deformation \cite{sreenivasan1991fractals}.
For \(\ell_{\mathrm{box}}<\ell_c\), the boxes increasingly sample the products of breakup, including local ligaments, detached droplets, and reconstructed VOF facets, whose geometry is closer to locally one-dimensional contours.
The crossover therefore reflects a finite-scale transition from a coarse-scale folding regime to a fine-scale fragment-and-contour regime.
Its location depends on the resolved interfacial morphology and numerical resolution, consistent with the known grid-scale sensitivity of atomization DNS~\cite{pairetti2020mesh,Kulkarni_Pairetti_Villiers_Popinet_Zaleski_2025}.

\begin{figure}[!t]
\centering
\includegraphics[width=0.95\columnwidth]{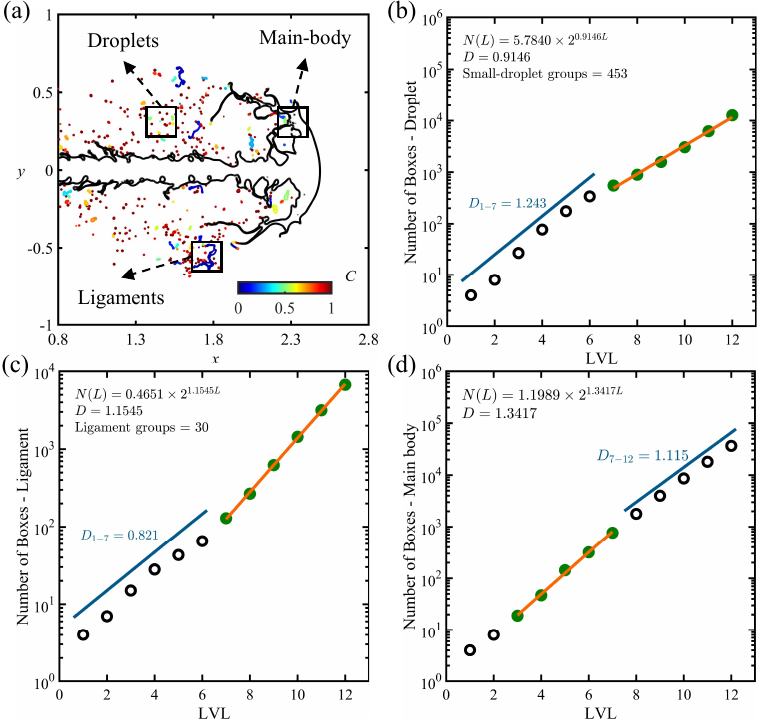}
\caption{
(a) Geometrical decomposition of the reference interface at maxlevel \(=12\) and \(u_{e,\max}=10^{-6}\). 
Detached components are colored by circularity \(C=4\pi A/P^2\).
We use $C=0.5$ to distinguish detached droplets ($C>0.5$) from ligament-like fragments ($C<0.5$).
Slender structures connected to the jet remain part of the main body.
(b)--(d) Box-counting results for droplets, detached ligament-like fragments, and the connected main body, respectively.
Green symbols indicate the subset-relevant fitting ranges, while blue guide fits show unused ranges.
Fine-scale fits characterize droplets and ligament-like fragments, whereas the connected main body is characterized by a coarse-scale fit.
}
\label{fig:fig4}
\end{figure}

This motivates a structure-resolved analysis of the interface.
We therefore decompose the maxlevel \(=12\), \(u_{e,\max}=10^{-6}\) interface into three geometrical subsets using both shape and connectivity.
Detached components are characterized by their circularity \(C=4\pi A/P^2\), where \(A\) and \(P\) are the  area and the perimeter of the component.
This compactness measure equals unity for a circle and decreases for elongated or irregular objects, making it a standard descriptor of particle and fragment shape \cite{wadell1935volume,blott2008particle}.
In Fig.~\ref{fig:fig4}(a), we use \(C=1/2\) as an operational threshold: detached components with \(C>1/2\) are classified as droplets, while detached components with \(C<1/2\) are classified as ligament-like fragments.
Slender structures that remain connected to the jet are not included in the detached ligament group, even if they are locally elongated; they are retained as part of the connected main body.
%This distinction follows the physical sequence of atomization, in which connected interfacial protrusions and ligaments precede detached fragments and droplets \cite{marmottant2004spray,villermaux2007fragmentation,eggers2008physics}.
Additional droplet-size and circularity statistics are provided in Sec.~S2 of the Supplemental Material~\cite{SupplementalMaterial}.

Figures~\ref{fig:fig4}(b)--\ref{fig:fig4}(d) show the box-counting results for the three subsets.
The green points indicate the fitting range used to extract the  effective dimension for the relevant subset, while the blue line fits show scale ranges that are not used for that subset.
This distinction is important because finite-range box-counting exponents depend on the scale interval over which the fit is performed \cite{theiler1990estimating}.
For detached droplets, the relevant range is the fine-scale one: at these scales, the boxes probe the local closed contours of nearly circular objects.
A coarse-scale fit would mainly reflect the sparse spatial distribution and separation of droplets, rather than the geometry of the droplet boundaries themselves.
For detached low-circularity fragments, the relevant range is also fine-scale, because these objects are filamentary and locally irregular.
Their effective dimension is therefore larger than that of nearly circular droplets, but still remains close to unity, the dimension for a smooth curve.
For the connected main body, the relevant range is instead the coarse-scale one.
This dimension measures the folded jet envelope and the connected protrusions that are still part of the primary liquid body.
Its fine-scale slope decreases toward unity,  the local linear segment limit, because, at those scales, the count increasingly follows reconstructed interface facets rather than the global organization of the jet.
The droplet value lies slightly below unity because it is a finite-range effective exponent measured for a sparse set of separated droplets, rather than the asymptotic boundary dimension of an isolated smooth closed contour.
To further distinguish droplet positions from their interfacial contours, we perform box counting on the droplet centroids (see Sec.~S3 of the Supplemental Material~\cite{SupplementalMaterial}).

Therefore, this subset dependence explains why the global dimension in Fig.~\ref{fig:fig2}(a) should be interpreted as an apparent full-interface descriptor rather than a unique fractal exponent.
The measured exponent depends on whether the covering boxes sample detached droplets, detached ligament-like fragments, or the connected envelope of the jet.
The physically relevant outcome is therefore a hierarchy of effective dimensions tied to both topology and observation scale, rather than a single dimension for the entire atomizing interface.

Additional comparisons between the VOF and CLSVOF interface representations are provided in Sec.~S4 of Supplemental Material~\cite{SupplementalMaterial}.

\begin{figure}[!t]
\centering
\includegraphics[width=0.75\columnwidth]{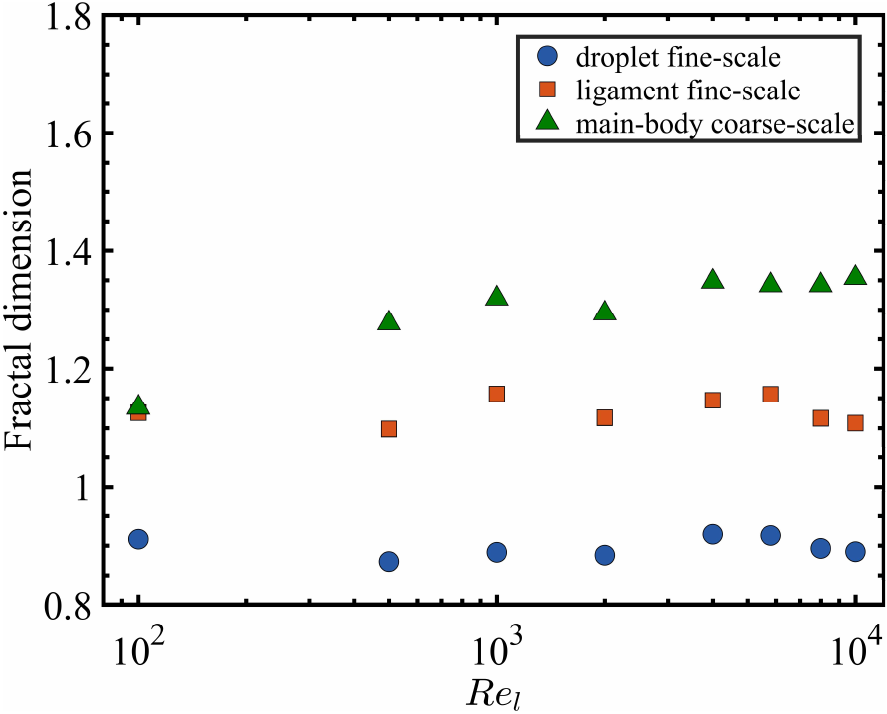}
\caption{
Reynolds-number dependence of the subset-relevant effective dimensions
at fixed \(We_g=200\).
The values are extracted using the fitting windows identified in
Fig.~\ref{fig:fig4}: fine-scale fits for detached droplets and
ligament-like fragments, and a coarse-scale fit for the connected
main body.
The figure is not intended as a scaling law for \(D(\Re_l)\), but as
a robustness test of the geometrical hierarchy.
Across \(\Re_l=100\)--\(10^4\), droplets have a very lacunar character, with dimension below one, ligament-like fragments have an intermediate
degree of lacunarity with $D_F\simeq 1.2$, and the connected main body retains the largest effective
dimension with $1.3 \le D_F\le  1.4$.
}
\label{fig:fig5}
\end{figure}

\textit{Reynolds-number robustness}--To assess the robustness of the subset-dependent hierarchy, the same analysis is repeated over the range $Re_l=100$ to $10^4$ at fixed \(We_g=200\).
Using fractal dimension as a geometrical response to a flow-control parameter is analogous to turbulent-interface and turbulent-flame studies, where measured dimensions have been compared across turbulence intensities to quantify interface or flame-front wrinkling \cite{sreenivasan1991fractals,gouldin1987application,mantzaras1989fractals,murayama1989fractal,thiesset2016geometrical}.
The main-body branch in Fig.~\ref{fig:fig5} is particularly similar in spirit to the compilations discussed by Sreenivasan \cite{sreenivasan1991fractals}: as the flow control parameter is varied, the dimension of the dominant connected interface provides a compact measure of geometrical wrinkling.
The comparison is methodological rather than quantitative.
The earlier turbulent-flame and turbulent-interface results generally concern three-dimensional fronts, surfaces, or their planar intersections, whereas the present main-body dimension is extracted from a 2D VOF-reconstructed curve over its coarse-scale fitting range.
The purpose here is therefore not to establish a scaling law for \(D(\Re_l)\), but to test whether the structure-dependent ordering identified above persists as the Reynolds number is varied.
Figure~\ref{fig:fig5} summarizes the subset-relevant effective dimensions extracted using the fitting windows identified in Fig.~\ref{fig:fig4}.

Across the explored range, the hierarchy remains intact.
Detached droplets stay below $D_F=1$, showing important lacunarity reminiscent of other point processes in 2D \cite{de1970case}, ligament-like fragments occupy an intermediate level, and the connected main body carries the largest coarse-scale effective dimension.
Increasing \(\Re_l\) changes the resolved vortical and interfacial organization, but does not collapse the hierarchy into a single exponent.
This persistence supports the interpretation of effective fractal dimension as a scale- and structure-resolved descriptor of the atomizing interface.

\textit{Conclusion}--The present 2D VOF-DNS of an atomizing liquid jet shows that the resolved liquid--gas interface should not be reduced to a single global fractal exponent.
Full-interface box-counting yields a useful apparent effective dimension, but this value masks a finite-scale crossover.
At coarse observation scales, the count is dominated by the folded envelope of the connected jet, where shear-layer vortices stretch and corrugate the liquid interface, consistent with the cascade view of turbulent interfacial deformation \cite{richardson1922weather,kolmogorov1962refinement,kolmogorov1991local,sreenivasan1991fractals}.
At finer observation scales, the count increasingly samples local ligaments, detached droplets, and reconstructed interface segments, so the measured dimension approaches that of locally one-dimensional contours.
The apparent dimension therefore depends on the observation scale and on the fitting window used to extract it, as expected for finite-resolution fractal estimates with both finite-size effects and small-scale cutoffs \cite{theiler1990estimating,falconer2013fractal,pairetti2020mesh,Kulkarni_Pairetti_Villiers_Popinet_Zaleski_2025}.

The structure-resolved analysis further shows that the relevant effective dimension depends on the geometrical subset being sampled.
Detached droplets are produced by a complex process\cite{eggers1997nonlinear,marmottant2004spray,villermaux2007fragmentation,eggers2008physics,Kulkarni_Pairetti_Villiers_Popinet_Zaleski_2025}. 
They exhibit strong lacunarity with $D_F < 1$. This is reminiscent of the strong lacunarity of star positions in the night star hypothesized by de Vaucouleurs \cite{de1970case}.
Detached ligament-like fragments occupy an intermediate scale, reflecting their elongated and locally irregular geometry.
The connected main body carries the largest coarse-scale dimension, which measures the global folding and spatial occupation of the jet envelope rather than the local roughness of individual interface segments.
The persistence of this ordering over \(\Re_l=100\)--\(10^4\) at fixed \(We_g=200\) indicates that this hierarchy is not limited to a single reference case.
Going forward, one would naturally like to extend this analysis to 3D atomizing jets \cite{Kulkarni_Pairetti_Villiers_Popinet_Zaleski_2025} and to explore oscillations in self-similarity caused for example by a hierarchical vortex structure \cite{smith1986lacunarity}.

\clearpage

\begin{acknowledgments}
We acknowledge St\'ephane Popinet for developing and maintaining the Basilisk code used in this work.
\end{acknowledgments}

% The \nocite command causes all entries in a bibliography to be printed out
% whether or not they are actually referenced in the text. This is appropriate
% for the sample file to show the different styles of references, but authors
% most likely will not want to use it.

\bibliographystyle{apsrev4-2}
\nocite{prasad1990measurement,
meneveau1990interface,iyer2020fractal,
ling2015multiscale,balachandar2020host,
sussman2000coupled,menard2007coupling,
herrmann2010detailed}

\bibliography{apssamp}% Produces the bibliography via BibTeX.

@article{Xu_Long_Wang_2023, 
title={Entrainment mechanism of turbulent synthetic jet flow}, 
volume={958}, 
DOI={10.1017/jfm.2023.102}, 
journal={J. Fluid Mech.}, 
author={Xu, Congyi and Long, Yanguang and Wang, Jinjun}, 
year={2023}, 
pages={A31}
}

@article{Soligo_Chiarini_Rosti_2025, 
title={Reynolds number effect on the flow statistics and turbulent–non-turbulent interface of a planar jet}, 
volume={1016}, 
DOI={10.1017/jfm.2025.10437}, 
journal={J. Fluid Mech.}, 
author={Soligo, Giovanni and Chiarini, Alessandro and Rosti, Marco Edoardo}, 
year={2025}, 
pages={A37}
}

@book{lefebvre2017atomization,
  title={Atomization and Sprays},
  author={Lefebvre, Arthur H. and McDonell, Vincent G.},
  year={2017},
  edition={2},
  publisher={CRC Press},
  url={https://books.google.fr/books?id=80OEDgAAQBAJ}
}

@book{bayvel2019liquid,
  title={Liquid atomization},
  author={Bayvel, Leoplod P},
  year={2019},
  url={https://books.google.fr/books?hl=zh-CN&lr=&id=JzmDDwAAQBAJ&oi=fnd&pg=PR7&dq=Liquid+atomization&ots=fog_Xw9TVP&sig=58Hmmor2zu4mfzLmERpGA0_iHOo&redir_esc=y#v=onepage&q=Liquid%20atomization&f=false},
  publisher={Routledge}
}

@article{pairetti2020mesh,
  title={Mesh resolution effects on primary atomization simulations},
  author={Pairetti, Cesar I and Damian, Santiago Marquez and Nigro, Norberto M and Popinet, Stephane and Zaleski, Stephane},
  journal={Atomization and Sprays},
  volume={30},
  number={12},
  year={2020},
  url={https://hal.science/hal-03080699/file/Preprint_MeshResolution2020.pdf},
  publisher={Begel House Inc.}
}

@article{Kulkarni_Pairetti_Villiers_Popinet_Zaleski_2025, 
title={The atomising pulsed jet}, 
volume={1009}, 
DOI={10.1017/jfm.2025.218}, 
journal={J. Fluid Mech.}, 
author={Kulkarni, Yash and Pairetti, Cesar and Villiers, Raphaël and Popinet, Stéphane and Zaleski, Stéphane}, 
year={2025}, 
pages={A35}
}

@article{hirt1981volume,
  title={Volume of fluid ({VOF}) method for the dynamics of free boundaries},
  author={Hirt, Cyril W and Nichols, Billy D},
  journal={J. Comput. Phys.},
  volume={39},
  number={1},
  pages={201--225},
  year={1981},
  url={https://d1wqtxts1xzle7.cloudfront.net/12246100/hirtnichols_surfer_jcp1981-libre.pdf?1390861259=&response-content-disposition=inline%3B+filename%3DVolume_of_Fluid_VOF_Method_for_the_Dynam.pdf&Expires=1789485992&Signature=JPaAyMNuuY2nCMEK~Rk2~n4z~l4Jh0oLktqQUQXpksn0syS~0D4-tnYJ8bNE059ktFXvjwQL9tv16RjFNhU2YpvceGuiC1Yr3n7lkmYi-a7kMVa6vwVeezVKdXiTldAAzh7tfFquBdV4EY5kE9PMM3oprwviSDrvrICXajRi1qQfshRchEsSf8BsPvMidkfzMQXEr-UWmXFlIuaH5Q9re88thtXOG0M7QsOBzN2-b4WqHsv-tzY3UaKMw2M7sjteDONzUPyU-lqUMjfi0DcJA3v5vI~HfEClkcDJJrCAFf5xQGcfCFWgrnNAOsdzFaQyKyUj52XXKOBgDjcPAjccSA__&Key-Pair-Id=APKAJLOHF5GGSLRBV4ZA},
  publisher={Elsevier}
}

@article{popinet2009accurate,
  title={An accurate adaptive solver for surface-tension-driven interfacial flows},
  author={Popinet, St{\'e}phane},
  journal={J. Comput. Phys.},
  volume={228},
  number={16},
  pages={5838--5866},
  year={2009},
  DOI={10.1016/j.jcp.2009.04.042}, 
  publisher={Elsevier}
}

@article{sussman2000coupled,
  title={A coupled level set and volume-of-fluid method for computing 3D and axisymmetric incompressible two-phase flows},
  author={Sussman, Mark and Puckett, Elbridge Gerry},
  journal={J. Comput. Phys.},
  volume={162},
  number={2},
  pages={301--337},
  year={2000},
  DOI={10.1006/jcph.2000.6537},
  publisher={Elsevier}
}

@article{menard2007coupling,
  title={Coupling level set/VOF/ghost fluid methods: Validation and application to 3D simulation of the primary break-up of a liquid jet},
  author={M{\'e}nard, Thibault and Tanguy, Sebastien and Berlemont, Alain},
  journal={Int. J. Multiphase Flow},
  volume={33},
  number={5},
  pages={510--524},
  year={2007},
  DOI={10.1016/j.ijmultiphaseflow.2006.11.001},
  publisher={Elsevier}
}

@article{herrmann2010detailed,
  title   = {Detailed Numerical Simulations of the Primary Atomization of a Turbulent Liquid Jet in Crossflow},
  author  = {Herrmann, Marcus},
  journal = {Journal of Engineering for Gas Turbines and Power},
  volume  = {132},
  number  = {6},
  pages   = {061506},
  year    = {2010},
  DOI={doi.org/10.1115/1.4000148},
}

@article{Sreenivasan_Meneveau_1986, 
   title={The fractal facets of turbulence},  
   volume={173}, 
   DOI={10.1017/S0022112086001209}, 
   journal={J. Fluid Mech.}, 
   author={Sreenivasan, K. R. and Meneveau, C.}, 
   year={1986}, 
   pages={357–386},
}

@book{richardson1922weather,
  title={Weather prediction by numerical process},
  author={Richardson, Lewis F},
  year={1922},
  url={https://books.google.fr/books?hl=zh-CN&lr=&id=D52d3_bbgg8C&oi=fnd&pg=PR11&dq=Weather+prediction+by+numerical+process&ots=62REg_5GTu&sig=ECQrKM1VxCA3JrxBzaIjxKcXbCg&redir_esc=y#v=onepage&q=Weather%20prediction%20by%20numerical%20process&f=false},
  publisher={Franklin Classics}
}

@article{kolmogorov1991local,
  title={The local structure of turbulence in incompressible viscous fluid for very large Reynolds numbers},
  author={Kolmogorov, Andrei Nikolaevich},
  journal={Proc. R. Soc. Lond. A},
  volume={434},
  number={1890},
  pages={9--13},
  year={1991},
  url={https://www.jstor.org/stable/51980?casa_token=nzwemWxgEuMAAAAA%3AXltWa_gKYYHvWWIqFLSBf9s007y-r8IEiWHCzIx0LVVy828eH04pqxIiS2AANEId-YMa2HYRqhmlhIPjDCuXQrZ1RR5YNFuk9CGUUGwQ6FntGH3-Gxqjag&seq=1}, 
  publisher={The Royal Society London}
}

@article{kolmogorov1962refinement,
  title={A refinement of previous hypotheses concerning the local structure of turbulence in a viscous incompressible fluid at high Reynolds number},
  author={Kolmogorov, Andrey Nikolaevich},
  journal={J. Fluid Mech.},
  volume={13},
  number={1},
  pages={82--85},
  year={1962},
  DOI={10.1017/S0022112062000518},
  publisher={Cambridge University Press}
}

@book{mandelbrot1983fractal,
  title={The fractal geometry of nature},
  author={Mandelbrot, Benoit B},
  publisher={W.H. Freeman and Co., New York},
  DOI={10.2307/2981858},
  year={1983}
}

@article{sreenivasan1991fractals,
  title={Fractals and multifractals in fluid turbulence},
  author={Sreenivasan, Katepalli R},
  journal={Annu. Rev. Fluid Mech.},
  volume={23},
  number={1},
  pages={539--604},
  url={https://users.ictp.it/~krs/pdf/1991_006.pdf},
  year={1991}
}

@article{constantin1991fractal,
  title={Fractal geometry of isoscalar surfaces in turbulence: theory and experiments},
  author={Constantin, Petre and Procaccia, Itamar and Sreenivasan, KR},
  journal={Phys. Rev. Lett.},
  volume={67},
  number={13},
  pages={1739},
  year={1991},
  DOI={10.1103/PhysRevLett.67.1739},
  publisher={APS}
}

@article{prasad1990quantitative,
  title={Quantitative three-dimensional imaging and the structure of passive scalar fields in fully turbulent flows},
  author={Prasad, Rahul R and Sreenivasan, KR},
  journal={J. Fluid Mech.},
  volume={216},
  pages={1--34},
  year={1990},
  DOI={10.1017/S0022112090000325},
  publisher={Cambridge University Press}
}

@article{gouldin1987application,
  title={An application of fractals to modeling premixed turbulent flames},
  author={Gouldin, Fl C},
  journal={Combustion and Flame},
  volume={68},
  number={3},
  pages={249--266},
  year={1987},
  DOI={10.1016/0010-2180(87)90003-4},
  publisher={Elsevier}
}

@article{el2022fractal,
  title={Fractal dimensions in fluid dynamics and their effects on the Rayleigh problem, the Burger's vortex and the Kelvin--Helmholtz instability},
  author={El-Nabulsi, Rami Ahmad and Anukool, Waranont},
  journal={Acta Mechanica},
  volume={233},
  number={1},
  pages={363--381},
  year={2022},
  DOI={10.1007/s00707-021-03128-9},
  publisher={Springer}
}

@article{muller1992implication,
  title={Implication of fractal geometry for fluid flow properties of sedimentary rocks},
  author={Muller, J and McCauley, JL},
  journal={Transport in Porous Media},
  volume={8},
  number={2},
  pages={133--147},
  year={1992},
  url={https://link.springer.com/article/10.1007/BF00617114},
  publisher={Springer}
}

@article{constantin1985determining,
  title={Determining modes and fractal dimension of turbulent flows},
  author={Constantin, Peter and Foias, Ciprian and Manley, Oscar P and Temam, Roger},
  journal={J. Fluid Mech.},
  volume={150},
  pages={427--440},
  year={1985},
  DOI={10.1017/S0022112085000209},
  publisher={Cambridge University Press}
}

@article{heinen2015classical,
  title={Classical liquids in fractal dimension},
  author={Heinen, Marco and Schnyder, Simon K and Brady, John F and L{\"o}wen, Hartmut},
  journal={Phys. Rev. Lett.},
  volume={115},
  number={9},
  pages={097801},
  year={2015},
  DOI={10.1103/PhysRevLett.115.097801},
  publisher={APS}
}

@article{el2023foam,
  title={Foam drainage equation in fractal dimensions: breaking and instabilities},
  author={El-Nabulsi, Rami Ahmad and Anukool, Waranont},
  journal={Euro. Phys. J. E},
  volume={46},
  number={11},
  pages={110},
  year={2023},
  DOI={10.1140/epje/s10189-023-00368-6},
  publisher={Springer}
}

@misc{basilisk,
  author       = {Popinet, St{\'e}phane and others},
  title        = {{Basilisk C}},
  howpublished = {\url{https://basilisk.fr/}},
  year         = {2013--2026},
  note         = {Last accessed: April 2026}
}

@article{popinet2015quadtree,
  title={A quadtree-adaptive multigrid solver for the Serre--Green--Naghdi equations},
  author={Popinet, St{\'e}phane},
  journal={J. Comput. Phys.},
  volume={302},
  pages={336--358},
  year={2015},
  DOI={10.1016/j.jcp.2015.09.009},
  publisher={Elsevier}
}

@article{wadell1935volume,
  title={Volume, shape, and roundness of quartz particles},
  author={Wadell, Hakon},
  journal={The Journal of geology},
  volume={43},
  number={3},
  pages={250--280},
  year={1935},
  DOI={10.1086/624298},
  publisher={University of Chicago Press}
}

@article{blott2008particle,
  title={Particle shape: a review and new methods of characterization and classification},
  author={Blott, Simon J and Pye, Kenneth},
  journal={Sedimentology},
  volume={55},
  number={1},
  pages={31--63},
  year={2008},
  DOI={10.1111/j.1365-3091.2007.00892.x},
  publisher={Wiley Online Library}
}

@article{marmottant2004spray,
  title={On spray formation},
  author={Marmottant, Philippe and Villermaux, Emmanuel},
  journal={J. Fluid Mech.},
  volume={498},
  pages={73--111},
  year={2004},
  DOI={10.1017/S0022112003006529},
  publisher={Cambridge University Press}
}

@article{villermaux2007fragmentation,
  title={Fragmentation},
  author={Villermaux, Emmanuel},
  journal={Annu. Rev. Fluid Mech.},
  volume={39},
  number={1},
  pages={419--446},
  year={2007},
  DOI={10.1146/annurev.fluid.39.050905.110214},
  publisher={Annual Reviews}
}

@article{eggers2008physics,
  title={Physics of liquid jets},
  author={Eggers, Jens and Villermaux, Emmanuel},
  journal={Rep. Prog. Phys.},
  volume={71},
  number={3},
  pages={036601},
  year={2008},
  DOI={10.1088/0034-4885/71/3/036601}
}

@article{theiler1990estimating,
  title={Estimating fractal dimension},
  author={Theiler, James},
  journal={Journal of the optical society of America A},
  volume={7},
  number={6},
  pages={1055--1073},
  year={1990},
  DOI={10.1364/JOSAA.7.001055},
  publisher={Optical Society of America}
}

@book{falconer2013fractal,
  title={Fractal geometry: mathematical foundations and applications},
  author={Falconer, Kenneth},
  year={2013},
  url={https://books.google.fr/books?hl=zh-CN&lr=&id=XJN7AgAAQBAJ&oi=fnd&pg=PA369&dq=Fractal+geometry:+mathematical+foundations+and+applications&ots=sK5ZcfyPeg&sig=56BUe53OsL9rS0yrfmtrKBJ0gTY&redir_esc=y#v=onepage&q=Fractal%20geometry%3A%20mathematical%20foundations%20and%20applications&f=false},
  publisher={John Wiley \& Sons}
}

@article{ling2015multiscale,
  title={Multiscale simulation of atomization with small droplets represented by a Lagrangian point-particle model},
  author={Ling, Yue and Zaleski, St{\'e}phane and Scardovelli, Ruben},
  journal={Int. J. Multiphase Flow},
  volume={76},
  pages={122--143},
  year={2015},
  DOI={10.1016/j.ijmultiphaseflow.2015.07.002},
  publisher={Elsevier}
}

@article{eggers1997nonlinear,
  title={Nonlinear dynamics and breakup of free-surface flows},
  author={Eggers, Jens},
  journal={Rev. Mod. Phys.},
  volume={69},
  number={3},
  pages={865},
  year={1997},
  DOI={10.1103/RevModPhys.69.865},
  publisher={APS}
}

@article{mantzaras1989fractals,
  title={Fractals and turbulent premixed engine flames},
  author={Mantzaras, J and Felton, PG and Bracco, FV},
  journal={Combustion and flame},
  volume={77},
  number={3-4},
  pages={295--310},
  year={1989},
  DOI={10.1016/0010-2180(89)90136-3},
  publisher={Elsevier}
}

@inproceedings{murayama1989fractal,
  title={Fractal-like character of flamelets in turbulent premixed combustion},
  author={Murayama, Motohide and Takeno, Tadao},
  booktitle={Symposium (International) on Combustion},
  volume={22},
  pages={551--559},
  year={1989},
  DOI={10.1016/S0082-0784(89)80062-1},
  organization={Elsevier}
}

@article{thiesset2016geometrical,
  title={Geometrical properties of turbulent premixed flames and other corrugated interfaces},
  author={Thiesset, F and Maurice, G and Halter, F and Mazellier, N and Chauveau, C and G{\"o}kalp, I},
  journal={Phys. Rev. E},
  volume={93},
  number={1},
  pages={013116},
  year={2016},
  url={https://hal.science/hal-01660272/file/Thiesset_PRE2016.pdf},
  publisher={APS}
}

@article{balachandar2020host,
  title={Host-to-host airborne transmission as a multiphase flow problem for science-based social distance guidelines},
  author={Balachandar, S and Zaleski, Stephane and Soldati, Alfredo and Ahmadi, Goodarz and Bourouiba, Lydia},
  journal={Int. J. Multiphase Flow},
  volume={132},
  pages={103439},
  year={2020},
  DOI={10.1016/j.ijmultiphaseflow.2020.103439},
  publisher={Elsevier}
}

@misc{SupplementalMaterial,
  note = {See Supplemental Material at [URL will be inserted by publisher] for material-interface fractal dimension and velocity-increment scaling, passive-scalar analyses and numerical sensitivity tests, droplet statistics, droplet-centroid box counting, and VOF/CLSVOF comparisons.},
  year=2026
}

@article{de1970case,
  title={The Case for a Hierarchical Cosmology: Recent observations indicate that hierarchical clustering is a basic factor in cosmology.},
  author={De Vaucouleurs, Gerard},
  journal={Science},
  volume={167},
  number={3922},
  pages={1203--1213},
  year={1970},
  DOI={ 10.1126/science.167.3922.1203},
  publisher={American Association for the Advancement of Science}
}

@article{smith1986lacunarity,
  title={Lacunarity and intermittency in fluid turbulence},
  author={Smith, Leonard A and Fournier, J-D and Spiegel, EA},
  journal={Physics Letters A},
  volume={114},
  number={8-9},
  pages={465--468},
  year={1986},
  DOI={10.1016/0375-9601(86)90695-X},
  publisher={Elsevier}
}

@article{villermaux2020fragmentation,
  title={Fragmentation versus cohesion},
  author={Villermaux, Emmanuel},
  journal={J. Fluid Mech.},
  volume={898},
  year={2020},
  DOI={doi.org/10.1017/jfm.2020.366},
  publisher={Cambridge University Press}
}

@article{le2008fractal,
  title={Fractal dimension and scale entropy applications in a spray},
  author={Le Moyne, Luis and Freire, V and {Queiros-Conde}, D },
  journal={Chaos, Solitons \& Fractals},
  volume={38},
  number={3},
  pages={696--704},
  year={2008},
  DOI={10.1016/j.chaos.2007.01.004},
  publisher={Elsevier}
}

@article{dumouchel2005experimental,
  title={Experimental analysis of liquid--gas interface at low Weber number: interface length and fractal dimension},
  author={Dumouchel, Christophe and Cousin, Jean and Triballier, Ka{\"e}lig},
  journal={Experiments in fluids},
  volume={39},
  number={4},
  pages={651--666},
  year={2005},
  DOI={10.1007/s00348-005-1005-5},
  publisher={Springer}
}

@article{grout2007fractal,
  title={Fractal analysis of atomizing liquid flows},
  author={Grout, S{\'e}bastien and Dumouchel, Christophe and Cousin, Jean and Nuglisch, Hans},
  journal={Int. J. Multiphase Flow},
  volume={33},
  number={9},
  pages={1023--1044},
  year={2007},
  DOI={10.1016/j.ijmultiphaseflow.2007.02.009},
  publisher={Elsevier}
}

@unpublished{rimbert2021direct,
    title={Direct and Inverse ``Cascade'' during Fragmentation of a Liquid Metal Jet into Water},
    author={Rimbert, Nicolas and Hadj-Achour, Miloud and Kewalramani, Gagan and Ji, Bowen and Labergue, Alexandre and Gradeck, Michel and Meignen, Renaud},
    year={2021},
    note={\href{https://hal.science/hal-03467196/}{preprint on HAL}}
}

@article{shavit1995fractal,
  title={Fractal dimensions of liquid jet interface under breakup},
  author={Shavit, Uri and Chigier, Norman},
  journal={Atomization and Sprays},
  volume={5},
  number={6},
  year={1995},
  DOI={10.1615/AtomizSpr.v5.i6.10},
  publisher={Begel House Inc.}
}

@article{chehroudi2004fractal,
  title={The fractal geometry of round turbulent cryogenic nitrogen jets at subcritical and supercritical pressures},
  author={Chehroudi, Bruce and Talley, Douglas},
  journal={Atomization and Sprays},
  volume={14},
  number={1},
  year={2004},
  DOI={10.1615/AtomizSpr.v14.i1.50},
  publisher={Begel House Inc.}
}

@article{castrejon2003fractal,
  title={The fractal dimension of an oil spray},
  author={Castrejon Garcia, R and Sarmiento Galan, A and Castrej{\'o}n Pita, JR and Castrej{\'o}n Pita, AA},
  journal={Fractals},
  volume={11},
  number={02},
  pages={155--161},
  year={2003},
  DOI={10.1142/S0218348X03001641},
  publisher={World Scientific}
}

@article{shavit1994fractal,
  title={Fractal Geometry of Liquid Jets under Breakup},
  author={Shavit, U. and Chigier, N.},
  journal={Fractals},
  volume={2},
  number={02},
  pages={277--281},
  year={1994},
  DOI={10.1142/S0218348X94000326},
  publisher={World Scientific}
}

@book{feder2013fractals,
  title={Fractals},
  author={Feder, Jens},
  year={2013},
  url={https://books.google.fr/books?hl=zh-CN&lr=&id=mgvyBwAAQBAJ&oi=fnd&pg=PA1&dq=Fractals&ots=drUMgYvcd8&sig=QYA6tTKEbUOR2gpSmpPtV988zl4&redir_esc=y#v=onepage&q=Fractals&f=false},
  publisher={Springer Science \& Business Media}
}

@Article{Herbert2026,
author={Herbert, Marco
and Hasslberger, Josef
and Bambauer, Maximilian
and Chakraborty, Nilanjan
and Klein, Markus},
title={Robustness and limitations of fractal-dimension estimation methods for complex fluid interfaces},
journal={Theoretical and Computational Fluid Dynamics},
year={2026},
volume={40},
pages={30},
doi={10.1007/s00162-026-00800-5},
url={https://doi.org/10.1007/s00162-026-00800-5}
}

@article{aniszewski2021parallel,
  title={Parallel, robust, interface simulator (PARIS)},
  author={Aniszewski, W and Arrufat, T and Crialesi-Esposito, M and Dabiri, S and Fuster, D and Ling, Y and Lu, J and Malan, L and Pal, S and Scardovelli, R and Tryggavson, G and Zaleski, S},
  journal={Computer Physics Communications},
  volume={263},
  pages={107849},
  year={2021},
  publisher={Elsevier}
}

@unpublished{2027-popinet,
  title={Multiphase},
  author={Popinet, S. and Zaleski, S.},
  note={submitted to J. Fluid Mech. Perspectives},
  year={2026}
}

@article{prasad1990measurement,
  title={The measurement and interpretation of fractal dimensions of the scalar interface in turbulent flows},
  author={Prasad, RR and Sreenivasan, KR},
  journal={Physics of Fluids A: Fluid Dynamics},
  volume={2},
  number={5},
  pages={792--807},
  year={1990},
  url={http://dx.doi.org/10.1063/1.857733},
  publisher={American Institute of Physics}
}

@article{meneveau1990interface,
  title={Interface dimension in intermittent turbulence},
  author={Meneveau, Charles and Sreenivasan, KR},
  journal={Physical Review A},
  volume={41},
  number={4},
  pages={2246},
  year={1990},
  url={https://doi.org/10.1103/PhysRevA.41.2246},
  publisher={APS}
}

@article{iyer2020fractal,
  title={Fractal iso-level sets in high-Reynolds-number scalar turbulence},
  author={Iyer, Kartik P and Schumacher, J{\"o}rg and Sreenivasan, Katepalli R and Yeung, Pui-Kuen},
  journal={Physical Review Fluids},
  volume={5},
  number={4},
  pages={044501},
  year={2020},
  url={https://doi.org/10.1103/PhysRevFluids.5.044501},
  publisher={APS}
}

% Supplemental Material -- original APS preprint typography.
\clearpage
\makeatletter
\@booleantrue\preprintsty@sw
\let\SMsavedsetup\setup@hook
\let\setup@hook\@empty
\input{aps12pt4-2.rtx}
\setup@hook
\let\setup@hook\SMsavedsetup
\let\section\section@preprintsty
\let\subsection\subsection@preprintsty
\let\subsubsection\subsubsection@preprintsty
\normalfont\normalsize
\ps@preprint
\makeatother
\onecolumngrid
\setcounter{figure}{0}
\setcounter{table}{0}
\setcounter{equation}{0}
\setcounter{section}{0}
\setcounter{secnumdepth}{1}
\renewcommand{\thefigure}{S\arabic{figure}}
\renewcommand{\thetable}{S\arabic{table}}
\renewcommand{\theequation}{S\arabic{equation}}
\renewcommand{\thesection}{S\arabic{section}}
\renewcommand{\theHfigure}{supp.\arabic{figure}}
\renewcommand{\theHtable}{supp.\arabic{table}}
\renewcommand{\theHequation}{supp.\arabic{equation}}
\renewcommand{\theHsection}{supp.\arabic{section}}

\begin{center}
{\large\bfseries Supplemental Material}
\end{center}

\section{Material-interface fractal dimension and velocity-incre\\
-ment scaling}
\label{sec:material-interface}

To investigate the physical origin of interfacial geometrical scaling, we consider a two-dimensional planar liquid jet entering a quiescent liquid pool, inspired by the dye-marked water-jet experiments of Prasad and Sreenivasan~\cite{prasad1990measurement,prasad1990quantitative}.
The goal is to demonstrate that multiscale stretching and folding by velocity fluctuations, a mechanism underlying the fractal geometry of passive-scalar interfaces in three-dimensional turbulence, also contributes to the development of geometrical scaling in two-dimensional simulations.
The material boundary between the injected and surrounding liquids is tracked using a non-diffusive volume-of-fluid (VOF) marker $f$, reconstructed by the piecewise-linear interface construction (PLIC) method.
The simulation is performed at $Re=4000$ until $t=100$, with a maximum refinement level of 12 and a velocity adaptation tolerance $u_{e,\max}=10^{-3}$.

Following Meneveau and Sreenivasan~\cite{meneveau1990interface}, the scale-dependent eddy viscosity is estimated from the characteristic velocity increment as
\begin{equation}
\nu_t(r)\sim r|\Delta u_r|
\sim u'r\left(\frac{r}{L}\right)^\xi,
\end{equation}
where $u'$ is the characteristic velocity fluctuation, $L$ the integral scale, and $\xi$ the velocity-increment exponent.
The intermittency correction relates $\xi$ to the generalized dissipation dimension $D_{1/3}$ through
\begin{equation}
\xi=\frac{1}{3}+\frac{2}{3}(3-D_{1/3}).
\end{equation}
The transport flux across a coarse-grained interface of area $S(r)$ then scales as
\begin{equation}
\Phi_{AB}(r)\sim
u'\Delta_{AB}S(r)\left(\frac{r}{L}\right)^\xi,
\end{equation}
where $\Delta_{AB}$ is the difference in the transported quantity across the interface.
Assuming scale-independent flux and $S(r)\sim r^{2-D_3}$ gives $D_3=2+\xi$.
For a two-dimensional interface with coarse-grained length $\mathcal{L}(r)\sim r^{1-D_f}$, the corresponding scaling argument yields
\begin{equation}
D_f=1+\zeta,
\end{equation}
where $\zeta$ denotes the measured velocity-increment exponent, corresponding to $\xi$ in the original three-dimensional formulation.

\begin{figure*}[h]
\centering
\includegraphics[width=0.85\textwidth]{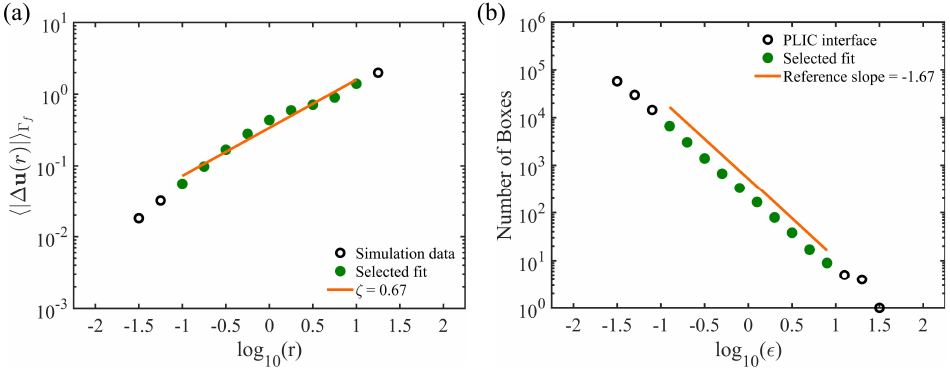}
\caption{
Material-interface fractal dimension and velocity-increment scaling at $t=100$.
(a) First-order velocity structure function near the material interface, yielding $\zeta\simeq0.67$.
(b) Box-counting results for the PLIC interface $f$ (black circles), with selected points (green) and the theoretical reference slope $-(1+\zeta)\simeq-1.67$ (orange).
}
\label{fig:supply1}
\end{figure*}

To allow sufficient time for the development of the jet, the snapshot at $t=100$ is selected for the following analysis.
The material boundary between the injected and surrounding liquids develops a highly convoluted geometry through successive stretching and folding by the flow.
Figure~\ref{fig:supply1}(a) presents the first-order velocity structure function $S_1(r)=\langle|\Delta\mathbf{u}(r)|\rangle$ evaluated near the material interface, exhibiting an approximate power-law scaling with $\zeta\simeq0.67$.
According to the relation $D_f=1+\zeta$ derived above, this gives a predicted dimension of approximately $1.67$.
The box-counting analysis of the PLIC-reconstructed interface yields an effective fractal dimension $D_f\simeq1.60$ over the selected scaling range [Fig.~\ref{fig:supply1}(b)].
The corresponding theoretical reference line with slope $-(1+\zeta)\simeq-1.67$ is superimposed on the box-counting data.
The proximity of the measured and predicted dimensions supports the connection between interfacial geometrical scaling and the multiscale velocity fluctuations responsible for material stretching and folding.
Although this agreement does not directly establish scale-independent transport, it is consistent with the proposed scaling argument over the finite range accessible to the simulation.

The geometrical features underlying this scaling can also be identified in the reconstructed interface shown in Fig.~\ref{fig:supply2}(a), where the injected liquid forms elongated and folded structures within the surrounding fluid.
Similar interfacial convolutions are visible in the experimental passive-scalar visualization of Prasad and Sreenivasan~\cite{prasad1990measurement}, reproduced in Fig.~\ref{fig:supply2}(b).
Although the experiment represents a planar section of a three-dimensional turbulent round jet, whereas our simulation is a genuinely two-dimensional planar flow, both exhibit complex interfacial geometries produced by fluid-induced stretching and folding.
Together with the velocity-increment scaling analysis, this qualitative correspondence supports the role of multiscale stretching and folding in the development of complex interfacial geometries in two-dimensional flows, without implying identical scaling behaviour in the two configurations.
The agreement is consistent with the proposed scaling argument over the finite range accessible to the simulation, without independently establishing scale-independent transport.
This comparison also supports the applicability of box counting for quantifying effective interfacial fractal dimensions in DNS.

\begin{figure*}[h]
\centering
\includegraphics[width=0.75\textwidth]{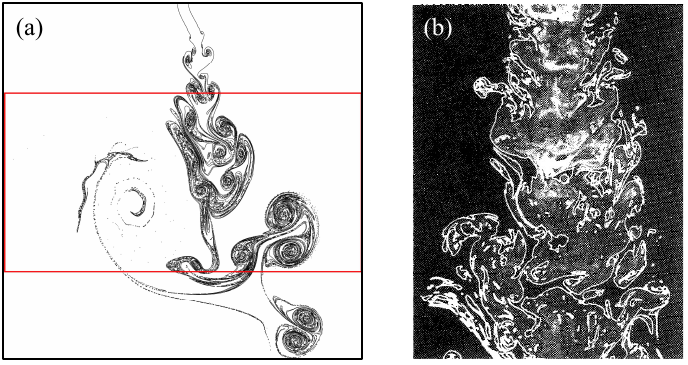}
\caption{
Simulated and experimental interface geometries.
(a) PLIC material interface at $t=100$, with the observation region extending from 8 to 24 inlet widths.
(b) Experimental passive-scalar interface in a turbulent round water jet, reproduced from Prasad and Sreenivasan~\cite{prasad1990measurement}.
}
\label{fig:supply2}
\end{figure*}

\clearpage

\section{Droplet statistics}
\label{sec:droplet-statistics}
Fig.~\ref{fig:supply7}(a) shows the number distribution of equivalent diameter \(d_{\mathrm{eq}}\) for different Maxlevel.
As the resolution is increased, the distribution extends toward smaller diameters, indicating that the small-droplet statistics remain sensitive to the finest resolved scale \cite{pairetti2020mesh,Kulkarni_Pairetti_Villiers_Popinet_Zaleski_2025,ling2015multiscale}.
In contrast, the large-droplet tail is comparatively less sensitive to resolution and remains roughly consistent with a Pareto-type power-law reference, \(N(d_{\mathrm{eq}})\sim d_{\mathrm{eq}}^{-2}\), over the resolved range \cite{balachandar2020host}.
The corresponding PDFs in Fig.~\ref{fig:supply7}(b) show the same sensitivity at small diameters, reinforcing that droplet-size statistics alone cannot define a unique resolution-independent breakup state.

The shape statistics clarify why the droplet subset identified in Fig.~4 of the main text is associated with a near-Euclidean fine-scale contour dimension.
Fig.~\ref{fig:supply7}(c) shows the circularity of detached components as a function of equivalent diameter.
Most detached objects above the grid scale cluster near \(C=1\), while lower circularity appears mainly for larger or more deformed fragments.
This is consistent with capillary breakup and relaxation, in which detached liquid fragments tend toward smooth closed contours in the present 2D setting, whereas recently detached or elongated fragments retain ligament-like shapes \cite{eggers1997nonlinear,marmottant2004spray,villermaux2007fragmentation,eggers2008physics}.
The circularity PDFs in Fig.~\ref{fig:supply7}(d) make this distinction quantitative: the number-based distribution is strongly peaked near \(C=1\), while the area-weighted distribution gives more importance to larger, less circular fragments.
Thus, the detached population is broad in size but dominated in number by nearly circular droplets, supporting the interpretation that the droplet effective dimension reflects local contour geometry rather than the coarse-scale folding of the connected jet.

\begin{figure*}[t]
\centering
\includegraphics[width=0.8\textwidth]{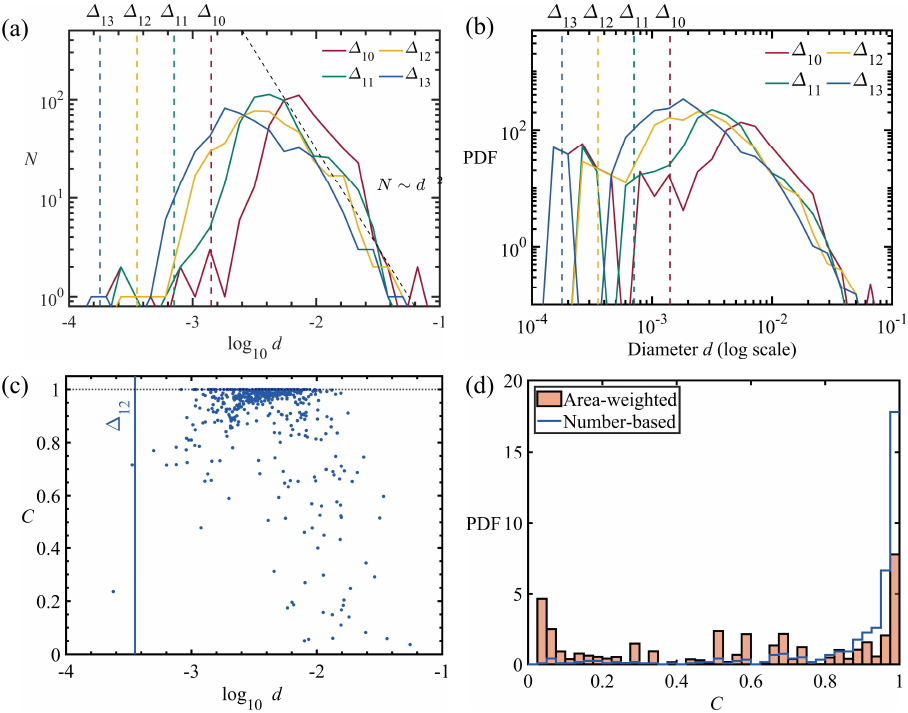}
\caption{
Droplet-size statistics and circularity of detached components.
(a) Number distribution of equivalent diameter \(d_{\mathrm{eq}}\) for different Maxlevel.
Increasing resolution reveals smaller droplets, while the large-droplet tail is compared with an approximate \(d_{\mathrm{eq}}^{-2}\) reference slope.
(b) Corresponding probability density functions of \(d_{\mathrm{eq}}\), showing the resolution dependence of the small-diameter range.
Vertical dashed lines indicate the grid scales \(\Delta_M\) for the corresponding Maxlevel.
(c) Circularity \(C\) of detached components as a function of equivalent diameter for the reference case with Maxlevel \(=12\).
The vertical line marks the grid scale \(\Delta_{12}\), and the horizontal dashed line indicates \(C=1\).
(d) Number-based and area-weighted PDFs of circularity.
The number-based PDF is dominated by nearly circular droplets with \(C\simeq1\), whereas area weighting gives more importance to larger, less circular fragments.
}
\label{fig:supply7}
\end{figure*}

\clearpage

\section{Spatial distribution of droplet centroids}
\label{sec:droplet-centroids}
To distinguish the spatial arrangement of droplets from the geometry of their boundaries, we replace each droplet in the subset analyzed in Fig.~4(b) of the main text by its two-dimensional area centroid.
The resulting set contains 453 points.
Each droplet contributes one point, irrespective of its size or shape, so the box count characterizes the spatial distribution of droplet positions rather than their interfacial contours.
Fig.~\ref{fig:supply8} shows these centroids as green points, revealing dense and sparse regions in their two-dimensional spatial distribution.
For clearer visualization, we overlay a grid corresponding to covering level \(L=5\) across the entire computational domain, showing that some boxes contain multiple centroids while others are empty.

\begin{figure*}[h]
\centering
\includegraphics[width=0.45\textwidth]{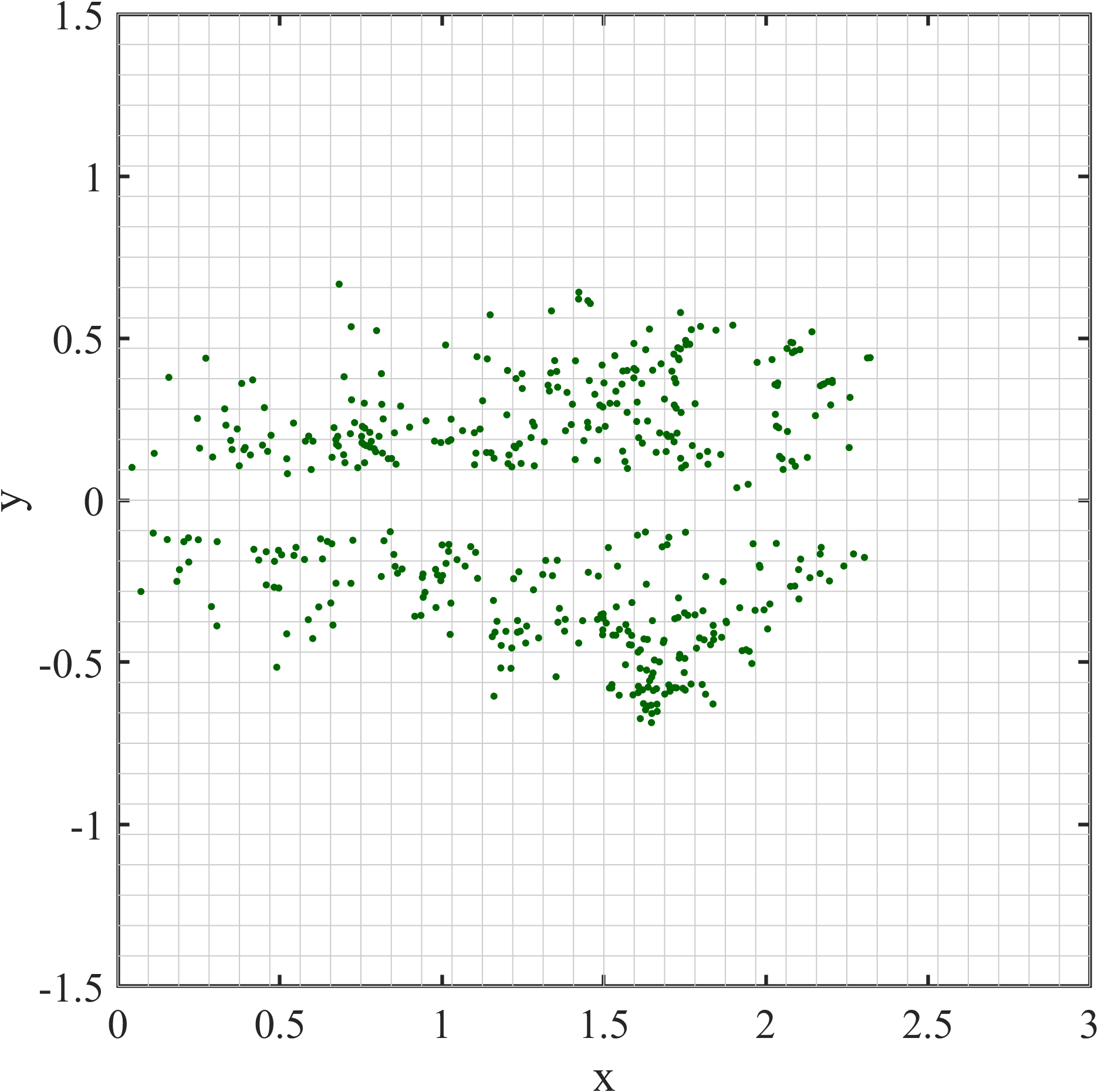}
\caption{
Spatial distribution of the 453 droplet area centroids, shown as green points.
The covering level \(L=5\) is chosen for clearer visualization of the centroid distribution within the overlaid grid.
}
\label{fig:supply8}
\end{figure*}

Fig.~\ref{fig:supply9} shows the number of boxes containing at least one centroid as the covering level is increased.
To assess whether the apparent scaling is sensitive to sparse sampling of the covering scales, Fig.~\ref{fig:supply9}(b) uses a finer scale spacing than Fig.~\ref{fig:supply9}(a).
The nearly unchanged fitted exponent supports the presence of an intermediate-scale scaling range.
Over an intermediate range of scales, the centroid set exhibits an effective box-counting exponent \(D_{\mathrm{cent}}\simeq1.31\).
This value is close to the coarse-scale effective dimension of the connected main body, \(D_{\mathrm{main}}\simeq1.34\), reported in Fig. 4(d) of the main text.
Thus, over their respective fitted ranges, the droplet positions and the connected jet body show comparable rates of growth in occupied-box count as the observation scale is refined.
This numerical proximity concerns their finite-scale spatial occupation.

At finer scales, the centroid box count approaches the total number of selected droplets and the effective slope tends toward zero.
This saturation occurs when distinct centroids are resolved in separate boxes.
It differs from the fine-scale counting of droplet contours, for which further refinement continues to resolve the boundaries of individual droplets.
The centroid analysis therefore separates the spatial organization of the detached droplet population from the local geometry of its interfaces.

\begin{figure*}[h]
\centering
\includegraphics[width=0.85\textwidth]{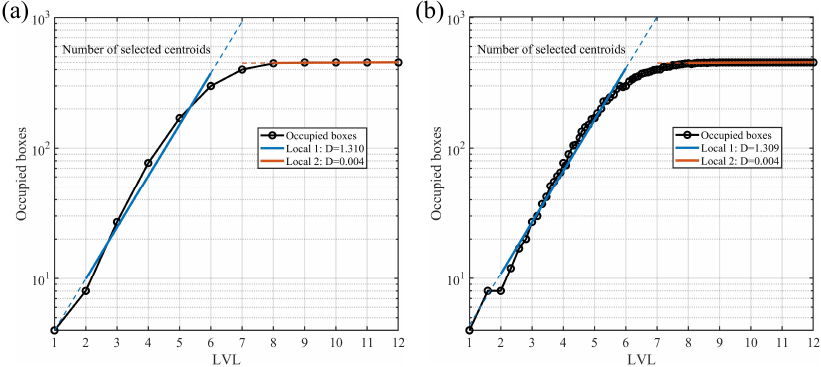}
\caption{
Box-counting analysis of the area centroids of the 453 droplets in the subset shown in Fig.~4(b) of the main text.
(a) Covering levels sampled at intervals of \(\Delta L=1\).
(b) Denser sampling with nominal intervals of \(\Delta L=0.2\).
Symbols show the number of boxes containing at least one centroid.
Blue lines indicate intermediate-scale fits, yielding effective exponents of 1.310 and 1.309 in (a) and (b), respectively, close to the coarse-scale main-body exponent \(D_{\mathrm{main}}\simeq1.34\).
Orange lines show the nearly horizontal fine-scale regime.
The horizontal dashed line marks the total number of selected centroids.
}
\label{fig:supply9}
\end{figure*}

\clearpage

\section{Comparison between VOF and CLSVOF}
\label{sec:vof-clsvof}
To assess whether the measured geometrical trends depend on the numerical representation of
the interface, we compare the geometrical volume-of-fluid (VOF) method used in the main text with a coupled level-set/volume-of-fluid
(CLSVOF) method. 
The VOF method advects the liquid volume fraction and reconstructs the interface geometrically  \cite{hirt1981volume,popinet2009accurate}, whereas CLSVOF supplements the volume-fraction field with a level-set representation of the
interface \cite{sussman2000coupled,menard2007coupling,herrmann2010detailed}. 
The physical parameters, computational domain, and refinement criteria are kept unchanged between the two calculations.

The preceding analysis showed that the measured dimension reflects the contributions of distinct interfacial structures, including the
connected jet body, ligaments, and detached droplets. 
Here, rather than separating these structures, we use the complete interface to
define a single aggregate geometrical diagnostic. 
In particular, the full-interface effective dimension is obtained from the slope of
\(\log_2 N(L)\) over the fixed intermediate-scale range \(3 \leq L \leq 7\), where \(N(L)\) is the number of occupied boxes at
box level \(L\). 
Thus, ``full interface'' refers to the inclusion of all interfacial components, not to a fit over all available box
levels. 
The same fitting range is used for every VOF and CLSVOF case, allowing us to test whether this intermediate-scale dimension is sensitive to the interface-capturing method.

Fig.~\ref{fig:supply10}(a) compares the full-interface effective dimensions obtained using VOF and CLSVOF as Maxlevel is varied at fixed \(u_{e,\max}=10^{-6}\). 
Both methods exhibit a similar increase from Maxlevel \(=8\) to \(9\), followed by nearly constant values at higher refinement levels.

A similar agreement is observed when varying the velocity-refinement tolerance at fixed Maxlevel \(=12\), as shown in Fig.~\ref{fig:supply10}(b). 
The two methods yield comparable effective dimensions and follow nearly identical trends with \(u_{e,\max}\). 
These results indicate that the intermediate-scale geometrical trends are relatively insensitive to the two interface representations considered here.

Fig.~\ref{fig:supply10}(c) and \ref{fig:supply10}(d) compare the droplet-size distributions obtained with the two methods for the
reference case. The VOF and CLSVOF calculations detect \(484\) and \(397\) detached components, respectively. Despite this difference in
the total number of detected components, the two methods produce similar distributions over the resolved large-droplet range. The
remaining differences occur mainly near the grid-sensitive small-diameter cutoff, where the formation and identification of
individual droplets are most strongly affected by the interface representation and the finest resolved scale.

\begin{figure*}[t]
\centering
\includegraphics[width=0.8\textwidth]{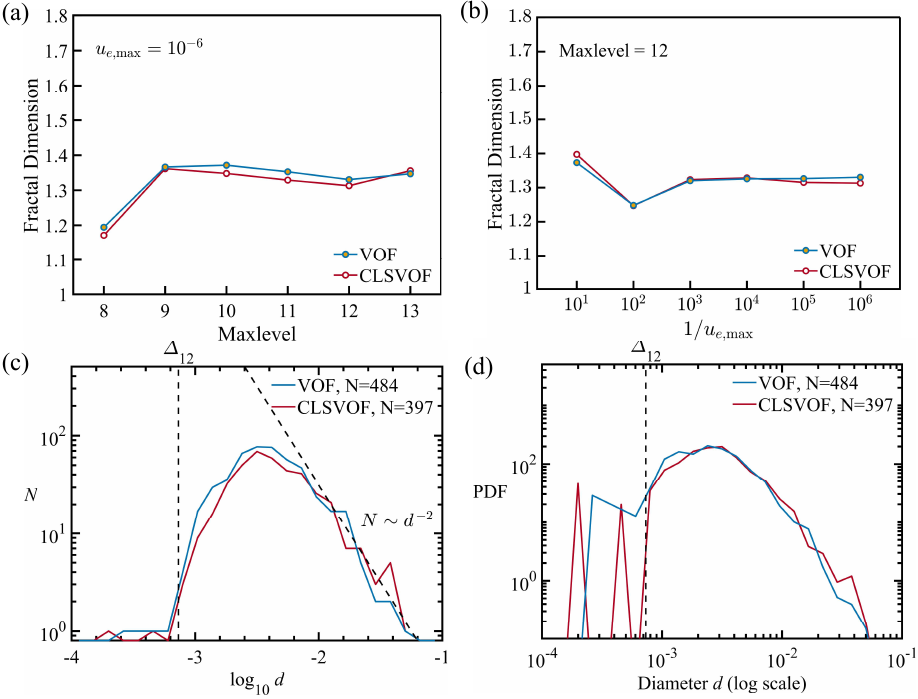}
\caption{
Comparison between the VOF and CLSVOF interface representations.
(a) Full-interface effective dimension as a function of Maxlevel at fixed \(u_{e,\max}=10^{-6}\).
(b) Full-interface effective dimension as a function of \(1/u_{e,\max}\) at fixed Maxlevel \(=12\).
(c) Number distributions of equivalent droplet diameter for the reference VOF and CLSVOF calculations; \(N\) denotes the number of
detected detached components.
(d) Corresponding probability density functions of equivalent diameter.
The vertical dashed lines in (c) and (d) indicate the finest grid scale \(\Delta_{12}\).
}
\label{fig:supply10}
\end{figure*}
\end{document}